\documentclass[a4paper,11pt]{article}
\pdfoutput=1 

\usepackage{jcappub}

\usepackage{graphicx}
\usepackage{dcolumn}
\usepackage{bm}
\usepackage{hyperref}

\newcommand{\mpl}{M_{\mathrm{Pl}}}
\newcommand{\mdm}{\mu}
\newcommand{\gsim}{\lower.7ex\hbox{$\;\stackrel{\textstyle>}{\sim}\;$}}
\newcommand{\lsim}{\lower.7ex\hbox{$\;\stackrel{\textstyle<}{\sim}\;$}}

\title{Black Hole Genesis of Dark Matter}

\author{Olivier Lennon,}
\emailAdd{olivier.lennon@physics.ox.ac.uk}

\author{John March-Russell,}
\emailAdd{jmr@thphys.ox.ac.uk}

\author{Rudin Petrossian-Byrne,}
\emailAdd{rudin.petrossian-byrne@physics.ox.ac.uk}

\author{and Hannah Tillim}
\emailAdd{hannah.tillim@physics.ox.ac.uk}

\affiliation{Rudolf Peierls Centre for Theoretical Physics, University of Oxford, 1 Keble Road, Oxford OX1
3NP, United Kingdom}

\abstract{We present a purely gravitational infra-red-calculable production mechanism for dark matter (DM).  The source of both
the DM relic abundance and the hot Standard Model (SM) plasma is a primordial density of
micro black holes (BHs), which evaporate via Hawking emission into both the dark and SM sectors.   The mechanism has four
qualitatively different regimes depending upon whether the BH evaporation is `fast' or `slow' relative to the initial Hubble rate, and whether the mass
of the DM particle is `light' or `heavy' compared to the initial BH temperature.  For each of these regimes we calculate the 
DM yield, $Y$, as a function of the initial state and DM mass and spin.  In the `slow' regime $Y$ depends on only the initial BH mass over a wide
range of initial conditions, including scenarios where the BHs are a small fraction of the initial energy density.  The DM is produced with a highly non-thermal energy spectrum, leading in the `light' DM mass regime ($\sim260\,\mathrm{eV}$ and above depending on DM spin) to a strong constraint from free-streaming, but also possible observational signatures in structure formation in the spin 3/2 and 2 cases.  The `heavy'  regime ($\sim1.2\times 10^8\,\mathrm{GeV}$ to $\mpl$ depending on spin) is free of these constraints and provides new possibilities for DM detection.   In all cases there is a dark radiation component predicted.}

\begin{document}
\maketitle
\flushbottom

\section{Introduction}
\label{sec:Intro}

It is now widely accepted that $\sim85\%$ of the matter content of the Universe is made up of a non-luminous component un-accounted for by Standard Model (SM) physics~\cite{Ade:2015xua}. While the nature and origin of this dark matter (DM) remains obscure, some properties are known -- a DM candidate must be cosmologically long-lived, cold (or possibly warm), with a dominant component(s) having mass(es) in the range: $10^{-22}\,\mathrm{eV}\lesssim \mdm\lesssim 10^{68}\,\mathrm{eV}$. The self- and DM-SM-interaction strengths are bounded above and consistent with being purely gravitational.

The most studied production mechanism for a relic abundance of particle-DM, thermal freeze-out~\cite{0038-5670-8-5-R07,Lee:1977ua}, assumes that the DM in the early Universe was in equilibrium with the hot SM plasma due to much-stronger-than-gravitational DM-SM interactions, but later fell out of equilibrium when annihilation rates failed to keep up with Hubble expansion. This mechanism depends on,
at minimum, the mass and couplings of the DM, and the number of degrees of freedom.  For suitably chosen values, often taken at the weak scale, freeze-out reproduces the relic abundance of the DM.  This mechanism, however, is now under strain from a combination of DM direct- and indirect-detection searches, and LHC bounds on new physics.  Thus it is of great importance to investigate other possible calculable production mechanisms that do not rely on close-to-weak-scale interactions
with the SM, such as freeze-in production of feebly-interacting DM from an initial state with negligible DM density~\cite{Hall:2009bx}.

Here we present a \emph{calculable, infra-red mechanism for DM production utilising only the mandatory coupling of DM to gravity}.  Specifically, we show
that both the DM abundance and the hot-Big-Bang SM plasma can result from an initial density of primordial black holes (pBHs).

It has long been thought that processes in the early Universe could have formed pBHs~\cite{Carr:1974nx}.
By Hawking's famous result~\cite{Hawking:1974sw}, these pBHs emit all states that couple to gravity and, if they are sufficiently small, can completely evaporate in the early Universe, contributing to the primordial radiation density.  As a subcomponent of this Hawking radiation will necessarily be the DM particles, pBHs can source some, or all, of the DM relic abundance. For earlier work in this direction, see~\cite{Matsas:1998zm, Bell:1998jk, Khlopov:2004tn, Fujita:2014hha}.

We will argue that a simple set of assumptions results in un-elaborate (and often largely initial-state independent) expressions for the final DM yield, $Y\equiv n_{\mathrm{DM}}/s_{\mathrm{tot}}$, and thus for $\Omega_{\mathrm{DM}} h^2$. (Here $s_{\mathrm{tot}}$ is the total SM entropy density of the Universe.)  Our starting hypotheses are: 1) There exists a population of micro pBHs with mean initial number density, $n_0$, with overlaid fluctuations.  We take the pBH masses to be narrowly peaked around an initial mass, $M_0$ (the generalisation of our results to somewhat broader initial mass distributions is straightforward).  Here we do \emph{not} seek to justify or provide a detailed mechanism for the existence of this population. 2) On large scales the initial energy density of BHs, $\rho_{\mathrm{BH}} = M_0 n_0$, inherits the approximately scale-invariant spectrum of density fluctuations, $\delta\rho/\rho\simeq 10^{-5}$, seen by cosmic microwave background and large-scale structure observations.  On tiny scales there can be significant departures from this fluctuation spectrum, e.g., due to Poisson variations in the number of BHs per initial Hubble patch. 3) Any remaining initial energy density not in the form of pBHs is in the form of a SM radiation component. Similarly to freeze-in production, we assume an initially negligible number density of DM particles. 4) The DM is in the form of cosmologically long-lived massive particles with
sub-Planckian mass and spin~$\leq 2$ which, further, have sufficiently small self- and DM-SM interactions that neither the freeze-out nor freeze-in processes are
operative (we will see this is often automatic as the DM mass giving the correct $\Omega_{\mathrm{DM}}h^2$ can be parametrically large).  
5) Stable Planckian-mass BH relics do not exist.
From these assumptions a successful mechanism of DM production follows, with noteworthy observational signatures in some cases.

In detail, in Section II we calculate the DM yield and SM `reheat' temperature, $T_{RH}$, at the end of pBH evaporation~\footnote{This is similar to a scenario discussed in~\cite{GarciaBellido:1996qt}, wherein the early Universe is reheated via Hawking emission of BHs post-inflation.}.  We present analytic expressions in the four qualitatively different regimes: `fast' or `slow' decay of the pBHs (with the general case being solved numerically) in both `light' and `heavy' DM scenarios. The observational constraints are applied in Section III, allowing us to find the acceptable parameter ranges of our mechanism. In Section IV, we consider variations of our basic mechanism with additional possible signatures and discuss future directions.

Finally, during manuscript preparation Ref.~\cite{Allahverdi:2017sks} appeared.  This work, although having some overlap,
differs considerably in both detail and general setup from ours.

\section{Dark Matter Yield}
\label{sec:EvappBH}

We first review the Hawking evaporation of a single, Schwarzschild BH, motivated by the fact that BHs with charge and angular momentum very quickly radiate these away (e.g.,~\cite{Davies:1978mf}).  The instantaneous temperature of a BH of mass $M$ is given by  $T_{\mathrm{BH}}=M^{2}_{\mathrm{Pl}}/(8\pi M)$ (we work in units where $\hbar=c=k_{\mathrm{B}}=1$) \cite{Hawking:1974sw}.
As the BH evaporates, $T_{\mathrm{BH}}$ scans all temperatures from an initial $T_0 = M^{2}_{\mathrm{Pl}}/(8\pi M_0)$ up to $T\sim M_{\mathrm{Pl}}$ where Hawking's
calculation breaks down.  In what follows we always assume that $T_0 >v_{\mathrm{EW}}=246$GeV,  corresponding to an initial mass of $M_{0}<10^{15}\mpl$. In this limit, all states in the SM are created ultra-relativistically throughout the BH's lifetime and so we need not consider mass thresholds or the physics of crossing through the weak or QCD scales.  Thus, during the entire evaporation process, the SM states can be well
approximated as massless.  (Our results generalise naturally to the case where $T_0$ is less than some SM masses.)  Taking, then, effectively massless
final states, the rate of BH mass-loss is~\cite{Page:1976df}
\begin{equation}
\frac{\mathrm{d}M}{\mathrm{d}t}=-\frac{\mpl^{4}}{M^{2}}\sum_{s,i}e_{s,i}g_{s,i}\equiv- e_{\mathrm{T}}\frac{\mpl^{4}}{M^{2}},
\label{eq:BHmassloss}
\end{equation}
where $i$ labels a state of spin $s$, $g_{s,i}$ is the associated number of degrees of freedom (dof), $e_{s,i}$ are the dimensionless grey-body `power-factors', given in Table~\ref{tbl:GreyBodyFactors}, and $ e_{\mathrm{T}}$ is the total emission coefficient, which does not depend upon the mass of the BH.  The grey-body factors appear due to a spin-dependent (and mass-dependent for particles of mass $m\not\ll T$) potential barrier outside of the horizon, which causes back-scattering into the BH.  These factors significantly alter the emission rates and must be taken into account for a correct evaluation of the DM yield.  For Hawking emission dominated by radiation into effectively massless SM states plus gravitons, $ e_{\mathrm{T}}\approx e_{\mathrm{T},\mathrm{SM}}\simeq 4.38\times 10^{-3}$. (This is a good approximation up to small corrections due the existence of the DM sector with small total dof $g_{\mathrm{DM}}\ll g_{\mathrm{SM}}\simeq 10^2$.) Thus the BH mass evolves as $M^{3}(t)=M_{0}^{3}-3 e_{\mathrm{T}} \mpl^{4}\left(t-t_{0}\right)$, where $M_{0}$ is the BH mass at initial time, $t_{0}$.

We consider two broad cases with regards to the production of DM particles by Hawking radiation: the first, where the initial temperature of the BH is greater than all mass scales--the `light' DM case--and the second, where it is below the DM mass, $\mdm$.  In the latter `heavy' case, creation of DM particles effectively only takes place after the BH has decayed sufficiently that $T\gsim\mdm$.  

Concretely, the emission rate of a species $i$ of spin $s$, and mass $\mu$, and with
energy in a range $(\omega,\omega+\text{d}\omega)$ is~\cite{Page:1976df}
\begin{equation}
\mathrm{d}\left(\frac{\mathrm{d}N_{s,i}}{\mathrm{d}t}\right)=\sum_{\ell,h} \frac{(2\ell+1)}{2\pi} \frac{\Gamma_{i,s,\ell,h}(\omega)}{\exp(\omega/T(t))+(-1)^{(2s+1)}} \mathrm{d}\omega\label{eq:emissionrate}
\end{equation}
where $\ell$ is the spherical harmonic, $h$ the helicity or polarisation of the emitted particle, and $\Gamma_{i,s,\ell,h}(\omega)$ is the absorption probability for that mode, which encodes the grey-body factor.   The total rate of emission, integrated over all final state energies $\omega=\sqrt{p^2 +\mu^2}$, with
$p\in [0,\infty]$, is, for our purposes, well approximated by
\begin{equation}
\frac{\mathrm{d}N_{s,i}}{\mathrm{d}t}\approx\frac{\mpl^{2}}{M}f_{s,i}g_{s,i} \Theta \left( d_s \frac{\mpl^2}{8\pi M} - \mu \right)
\label{eq:approxemissionrate}
\end{equation}
where the coefficients $f_{s,i}$ are dimensionless greybody `rate-factors', with values given in Table~\ref{tbl:GreyBodyFactors}, $\Theta$ is
the Heaviside step-function, and $d_s$ is a spin-dependent dimensionless coefficient we have found from numerical and semi-analytic
solutions of the exact Hawking emission equations.  To a sufficient approximation $d_{\mathrm{bosons}}\simeq 3.2$, while $d_{\mathrm{fermions}} \simeq
3.6$ (this simple parametrisation ignores some sub-leading residual dependence on spin, $s=0,1,2$ in the boson case, and $s=1/2,3/2$ in the fermion case, arising from the differing energy dependence of the grey-body factors).  Eqs.~\eqref{eq:BHmassloss}~and~\eqref{eq:approxemissionrate} imply that the total number of particles, per species, emitted over the BH's life is
\begin{equation}
N_{s,i}\simeq\frac{f_{s,i}g_{s,i}}{2 e_{\mathrm{T}}}\left(\frac{M_{0}}{\mpl}\right)^{2}
\left\{
\begin{matrix}
1 &(T_{0}>\mu/d_{s})\\
d_{s}^2 T_{0}^{2}/\mu^{2} & (T_{0}<\mu/d_{s})
\end{matrix}
\right. .
\label{eq:particlenumber}
\end{equation}

\begin{table}[t]\centering
	\begin{tabular}{| c | c | c |}
	\hline
		Spin & $e_{s}$ & $f_{s}$ \\
		\hline
		$0$ 	& $7.24\times10^{-5}$ & $6.66\times10^{-4}$ \\
		$1/2$ & $4.09\times10^{-5}$ & $2.43\times10^{-4}$ \\
		$1$	& $1.68\times10^{-5}$ & $7.40\times10^{-5}$ \\
		$3/2$ & $5.5\times10^{-6}$ & $2.1\times10^{-5}$ \\
		$2$ & $1.92\times10^{-6}$ & $5.53\times10^{-6}$ \\
		\hline
	\end{tabular}
	\caption{The numerical power, $e_s$, and rate, $f_s$, greybody factors per degree of freedom for neutral particles of given spin $s$, extracted from~\cite{Page:1976df, MacGibbon:1990zk,Carr:2016hva}. These have been summed over the dominant angular momentum emission modes~\cite{Sanchez:1977si}. For spin $3/2$ we have estimated the coefficients. There is a small, sub-dominant correction to  $e_s$ and $f_s$ if particles are charged.}
	\label{tbl:GreyBodyFactors}
\end{table}

{\bf Light Dark Matter:} Consider, first, the case where the DM mass is less than the initial effective BH temperature $\mdm<d_{s} T_0$.  The evolution equations of the energy densities of the pBHs, $\rho_{\mathrm{BH}}=M(t)n(t)$, and the SM radiation, $\rho_{\mathrm{rad}}$, in the early Universe are:
\begin{align}
\frac{\mathrm{d}\rho_{\mathrm{BH}}}{\mathrm{d}t}&+3H\rho_{\mathrm{BH}}=- e_{\mathrm{T}}\frac{\mpl^{4}}{M^{2}}n~;\label{eq:BHErgDensity}\\
\frac{\mathrm{d}\rho_{\mathrm{rad}}}{\mathrm{d}t}&+4H\rho_{\mathrm{rad}}=+ e_{\mathrm{T}}\frac{\mpl^{4}}{M^{2}}n~.\label{eq:RadErgDensity}
\end{align}
We solve these equations together with Eq.~\eqref{eq:BHmassloss} and the Friedmann equation, $H(t)^2 = 8\pi(\rho_{\mathrm{BH}}+\rho_{\mathrm{rad}})/3\mpl^2$, 
subject to $M_0>\mpl$ and $\rho_{\mathrm{BH}}(0)=M_0 n_0<\mpl^{4}$. We also assume that the process begins in a pBH-matter-dominated universe
with $\rho_{\mathrm{rad}}(0)\approx0$ -- in Section~IV we discuss the (in)sensitivity of our results to the relaxation of this assumption.
It should be noted that a number of studies into formation process for pBHs give the number of BHs per initial Hubble patch, $4\pi n_{0}/3H_{0}^{3}\sim1$ (e.g.,~\cite{Carr:1974nx, GarciaBellido:1996qt, Hawking:1987bn, Polnarev:1988dh}), though we do not impose this constraint here.

The resulting solution is well described by one of two analytic limits, here denoted as `fast' and `slow' decay.  These limits are characterised by the ratio of timescales of the system: the characteristic BH decay time, $\tau_{\mathrm{dec}}(M(t))$, and the Hubble time, $t_{H}(t)$. We define $B(t)\equiv t_{H}/\tau_{\mathrm{dec}}$, with initial value
\begin{equation}
(B_{0})^{2}= e_{\mathrm{T}}^{2}\frac{3}{2\pi}\frac{\mpl^{3}}{n_{0}}\left(\frac{\mpl}{M_{0}}\right)^{7}.
\label{eq:B0def}
\end{equation}
When $B_{0}\gg1$ we have fast decay, while for $B_{0}\ll1$ the decay is initially slow. Note that all decays become fast towards the end of the BH lifetime.

The SM interactions are sufficiently strong that $\rho_{\mathrm{rad}}$ thermalises quickly, while, as stated in the introduction (assumption 4), the DM does not interact significantly with either itself or the SM, so that no other processes affect the DM yield.  The two important calculable quantities from these analytic cases are the temperature of the SM radiation bath at the end pBH decay -- the `reheat' temperature, $T_{\mathrm{RH}}$ -- and the yield of the DM.  We will minimally impose that $T_{\mathrm{RH}}>3\,\mathrm{MeV}$, the starting temperature of Big Bang Nucleosynthesis (BBN)~\cite{Kawasaki:2000en,Jedamzik:2009uy}, and that the DM yield and mass satisfy
$\mdm Y=0.43\,\mathrm{eV}$, ensuring, respectively, that BBN is standard and $\Omega_{\mathrm{DM}}h^{2}=0.11$ today, as observed. 

From our analytic solution we find, in the case of slow initial decay, $B_{0}\ll1$,
\begin{equation}
T_{\mathrm{RH}}^{\mathrm{slow}}\simeq1.09\frac{ e_{\mathrm{T}}^{1/2}}{g_{*}^{1/4}}\mpl\left(\frac{\mpl}{M_{0}}\right)^{3/2},
\label{eq:ReheatTempSlow}
\end{equation}
where $g_{*}$ is the effective number of SM dof at $T_{\mathrm{RH}}$. From the ensuing total entropy density, $s_{\mathrm{tot}}$ (well approximated
by that in the SM radiation bath in our limit $g_{\mathrm{DM}}\ll g_{\mathrm{SM}}$), and using Eq.~\eqref{eq:particlenumber}, the yield of
species $i$ with spin $s$ at the end of the decay process is then given by
\begin{equation}
\label{eq:YieldSlow}
Y^{\mathrm{slow}}\equiv\frac{n_{s,i}}{s_{\mathrm{tot}}}\simeq0.49\frac{f_{s,i}g_{s,i}}{g_{*}^{1/4} e_{\mathrm{T}}^{1/2}}\left(\frac{\mpl}{M_{0}}\right)^{1/2}.
\end{equation}
Notice that both of these expressions are \emph{independent} of the initial pBH number density. 

In the fast case, $B_0 \gg 1$, the pBHs effectively dump all of their energy into radiation instantaneously without significant Hubble red-shifting.  Thus, 
\begin{equation}
\label{eq:ReheatTempFast}
T_{\mathrm{RH}}^{\mathrm{fast}}\simeq\left(\frac{30n_{0}M_{0}}{\pi^{2}g_{*}}\right)^{1/4},
\end{equation}
with corresponding particle yield
\begin{equation}
\label{eq:YieldFast}
Y^{\mathrm{fast}}\simeq0.50\frac{f_{s,i}g_{s,i}}{g_{*}^{1/4} e_{\mathrm{T}}}\left(\frac{n_{0}}{\mpl^{3}}\right)^{1/4}\left(\frac{M_{0}}{\mpl}\right)^{5/4}.
\end{equation}

\begin{figure}
\centering
  \includegraphics[scale=0.70]{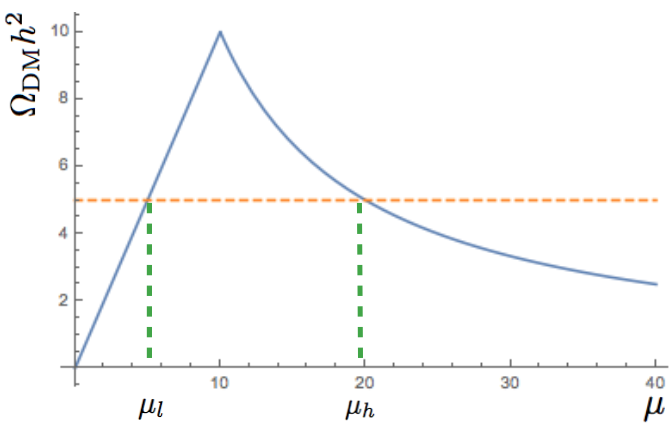}
\caption{\label{fig:Ymu}
	Schematic behaviour of $\Omega_{\mathrm{DM}}h^{2}$ as a function of DM mass $\mdm$, here given in arbitrary units, with initial pBH properties $M_0$ and $n_0$ held fixed. The orange dashed line represents the observed value of $\Omega_{\mathrm{DM}}h^{2}$. The green dashed lines denote the `light' ($\mdm_{l}$) and `heavy' ($\mdm_{h}$) DM masses that are consistent with the observed relic abundance.} 
\end{figure}

{\bf Heavy Dark Matter:} We now consider the case $\mdm>d_{s} T_0$. The reheat temperature remains, in both slow and fast cases, of the form given in Eqs.~\eqref{eq:ReheatTempSlow}~and~\eqref{eq:ReheatTempFast}, up to small corrections of $\mathcal{O}(g_{\mathrm{DM}}/g_{\mathrm{SM}})$. From Eq.~\eqref{eq:particlenumber}, we can then write the yield in this scenario as
\begin{equation}
Y_{h}=Y_{l}\frac{T_{0}^{2}}{\mdm_{h}^{2}}d_{s}^{2},
\end{equation}
where the subscripts $h$ and $l$ refer, respectively, to the heavy and light cases.
In Fig.~\ref{fig:Ymu}, we see that the observable quantity, $\mdm Y$ (equivalently $\Omega_{\mathrm{DM}}h^{2}$), exhibits two contrasting behaviours as a function of the DM mass $\mdm$,  namely linearly rising for small $\mdm$, while decreasing as $1/\mdm$ for large values.  Thus, for any value of $\mdm Y$ lower than the maximum (we find that $\mdm Y=0.43\,\mathrm{eV}$ always satisfies this), \emph{two} solutions for $\mdm$ exist -- one in each regime.  The two solutions satisfy $\mdm_{h}\mdm_{l} = T_{0}^{2}d_{s}^{2}$.

\begin{figure*}[t]\centering
	\begin{minipage}{\columnwidth}
		\includegraphics[width=0.47\columnwidth]{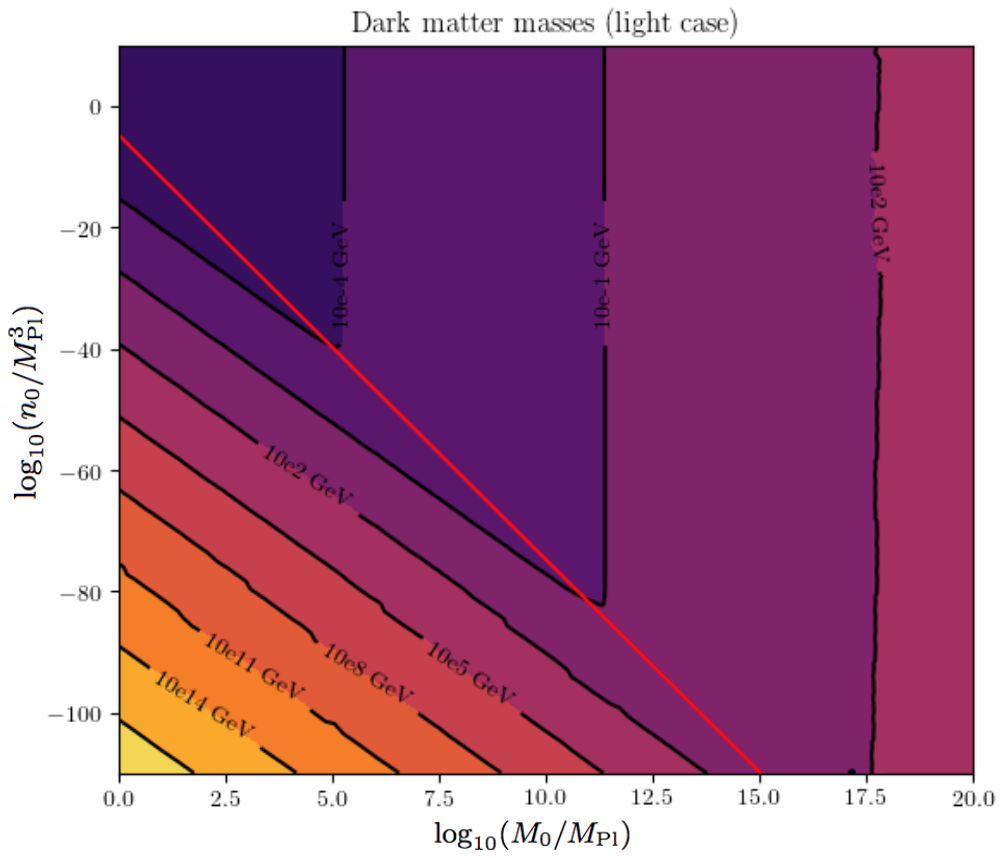}
		\hspace{0.5cm}
		\includegraphics[width=0.47\columnwidth]{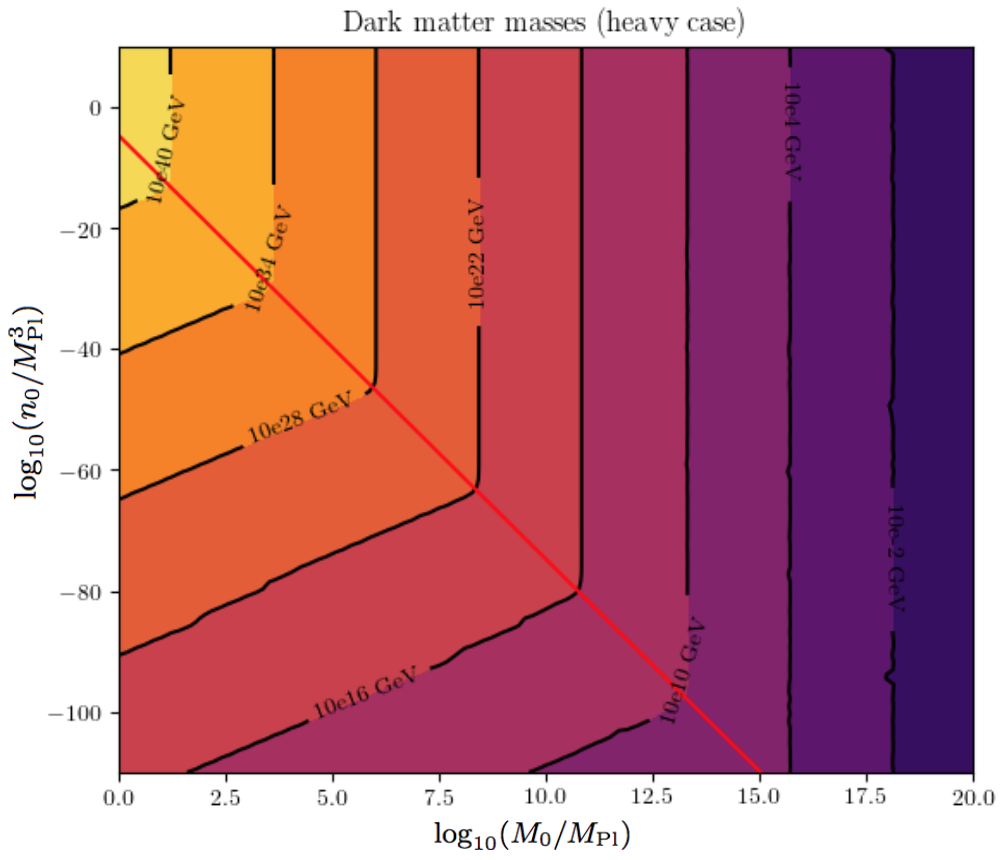}
	\end{minipage}
	\caption{
Contours of DM mass $\mdm$ giving observed $\Omega_{\mathrm{DM}}h^{2}$,  for the `light' and `heavy' DM cases, as a function of $\log_{10}(M_0/\mpl)$ and $\log_{10}(n_0/\mpl^3)$ across the \emph{a priori} allowed parameter space \emph{before constraints are imposed}.  The red line is $B_0 = 1$.}
	\label{fig:mass_plots}
\end{figure*}

{\bf Numerics:} We have numerically solved the full set of coupled equations (Eqs.~\eqref{eq:BHmassloss},~\eqref{eq:BHErgDensity},~\eqref{eq:RadErgDensity}, and the Friedmann equation). Our results, shown in Fig.~\ref{fig:mass_plots}, confirm our limiting approximations are 
excellent over the majority of parameter space.  The line $B_0 = 1$ is also confirmed to be a good discriminant, marking a quick but smooth transition between the two regimes. The plots also make obvious the different behaviour of $\mdm_l$ and $\mdm_h$ over the parameter space, i.e. whether they increase or decrease towards the origin in the upper left corner where $M_0$ is small and $n_0$ large.

\begin{figure*}[t]
	\begin{minipage}{\columnwidth}
		\includegraphics[width=0.47\columnwidth]{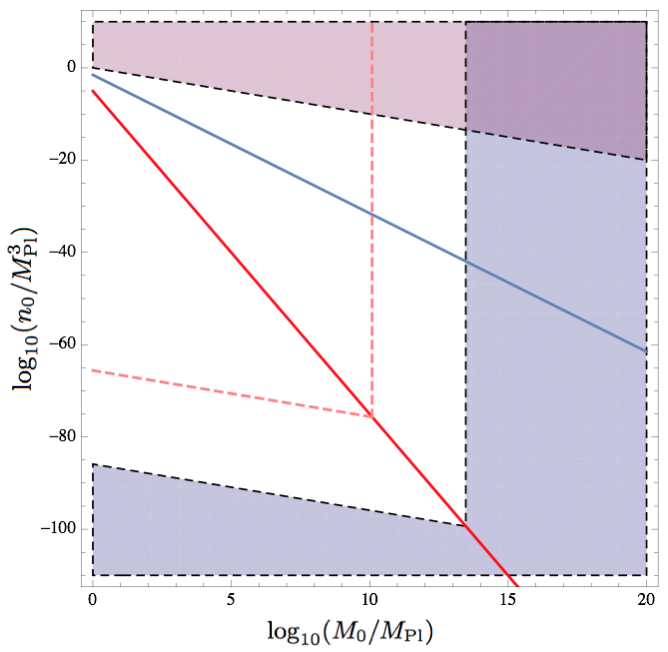}
		\hspace{0.5cm}
		\includegraphics[width=0.47\columnwidth]{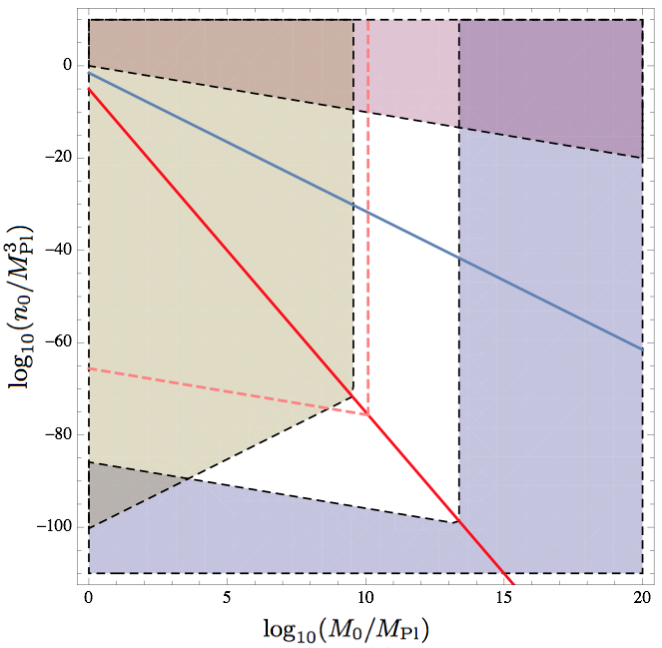}
	\end{minipage}
	\caption{
Constraints in the $\log_{10}(M_0/\mpl)$--$\log_{10}(n_0/\mpl^3)$ plane for `light' and `heavy' DM cases (left and right panels), for a single real 
$s=0$ dof (other spins display precisely the same features, differing only in numerical values)
\emph{before} the additional, strong constraint coming from free-streaming is imposed.  (In the case of `light' DM this strongly
DM-spin-dependent constraint excludes DM spin $s\leq 1$, marginally excludes $s=3/2$, and allows a region of the spin $s=2$ case to survive.
There are no significant constraints on the `heavy' DM case from free-streaming.  See text for more details.)   
The red line ($B_0 = 1$) marks the boundary between the `fast' and `slow' regimes; the blue line corresponds to one pBH per initial Hubble patch
with increasing pBH number density below.  The blue shaded regions are disallowed as $T_{RH}<3\,\mathrm{MeV}$.  The pink, dotted line is $T_{RH}=200\,\mathrm{GeV}$. 
The upper purple, and in the `heavy' case khaki, shaded regions are excluded as $\rho_{BH}(0)>\mpl^4$ and $\mdm>\mpl$, respectively.}
	\label{fig:constraint_plots}
\end{figure*}

\section{Observational Constraints and Possible Signatures}

In order for our model to follow standard cosmological results from BBN onwards, we minimally require that $T_{\mathrm{RH}}>3\,\mathrm{MeV}$ \cite{Kawasaki:2000en,Jedamzik:2009uy}.  This constraint, along with $M_0>\mpl$ and $\rho_{\mathrm{BH}}(0)=M_0 n_0<\mpl^{4}$, is shown in Fig.~\ref{fig:constraint_plots} in the `light' DM mass (left) and `heavy' DM mass (right) cases.  We show only constraint plots for a single real $s=0$ dof as other spins
display precisely the same features, differing only in numerical values.  These, together with the DM mass contour plots given in Fig.~\ref{fig:mass_plots}, lead to the maximally allowed ranges for the DM mass in Table~\ref{tbl:DMMassRanges} in each of the four regimes, and for DM spins $0\leq s \leq 2$  (these are calculated \emph{before} imposition of the additional, strong constraint coming from free-streaming).

\begin{table*}[t]\centering
\scalebox{0.85}{
	\begin{tabular}{| c | c | c | c | c | c |}
	\hline
		Spin & $g_{s}$ & $\mdm/\mathrm{GeV}$ (slow, light) & $\mdm/\mathrm{GeV}$ (slow, heavy) & $\mdm/\mathrm{GeV}$ (fast, light) & $\mdm/\mathrm{GeV}$ (fast, heavy) \\
		\hline
		$0$ 	& $1$ & $\left[2.6\times10^{-7},0.80\right]$ & $\left[3.4\times10^{9},\mpl\right]$ & $\left[3.1\times10^{-7},2.8\times10^{13}\right]$ & $\left[2.9\times10^{9},\mpl\right]$ \\
		$1/2$ & $2$ & $\left[3.6\times10^{-7},1.1\right]$ & $\left[3.1\times10^{9},\mpl\right]$ & $\left[4.2\times10^{-7},3.9\times10^{13}\right]$ & $\left[2.6\times10^{9},\mpl\right]$ \\
		$1$	& $3$ & $\left[7.8\times10^{-7},2.4\right]$ & $\left[1.1\times10^{9},\mpl\right]$ & $\left[9.2\times10^{-7},8.5\times10^{13}\right]$ & $\left[9.6\times10^{8},\mpl\right]$ \\
		$3/2$ & $4$ & $\left[2\times10^{-6},6\right]$ & $\left[5\times10^{8},\mpl\right]$ & $\left[2\times10^{-6},2\times10^{14}\right]$ & $\left[5\times10^{8},\mpl\right]$ \\
		$2$ & $5$ & $\left[6.3\times10^{-6},19\right]$ & $\left[1.4\times10^{8},\mpl\right]$ & $\left[7.4\times10^{-6},6.8\times10^{14}\right]$ & $\left[1.2\times10^{8},\mpl\right]$ \\
		\hline
	\end{tabular}}
	\caption{Limits on DM mass giving the correct relic abundance in the four regimes of `light'  and `heavy' DM mass and `slow' and `fast' BH decay, which satisfy our observational and theoretical constraints (except free-streaming in `light' case -- see text).  A single DM particle of given spin is assumed. In particular, given the large masses allowed above, this mechanism provides a new way for producing so-called `WIMPzillas'~\cite{Kolb:1998ki} in the early Universe.}
	\label{tbl:DMMassRanges}
\end{table*}

In the slow regime, $T_{\mathrm{RH}}$ and $Y$ (Eqs.~\eqref{eq:ReheatTempSlow}~and~\eqref{eq:YieldSlow}) are independent of $n_{0}$, and iso-contours of $T_{\mathrm{RH}}$, $\mdm$ giving the observed $\Omega_{DM}h^2$ and $M_{0}$ will all be parallel.  Thus, constraining the reheat temperature and the initial BH mass will provide constraints on the DM mass.  Since, in the light (respectively, heavy) case, the DM mass contours increase (decrease) with increasing $M_{0}$, we find that $T_{\mathrm{RH}}\lesssim3\,\mathrm{MeV}$ sets an upper (lower) bound on $\mdm$, while $M_{0}>\mpl$ sets a lower (upper) bound. However, in the heavy case, we must exclude the region where $\mdm>\mpl$.  This then leads to an \emph{upper bound on the reheat temperature in the heavy case} ranging from $T_{\mathrm{RH}}\lesssim1.3\,\mathrm{TeV}$ for $s=0$ up to $T_{\mathrm{RH}}\lesssim8.5\,\mathrm{TeV}$ for $s=2$.

In the fast regime, the mass contours are of differing gradients for the light and heavy cases, setting different locations for the bounds on the DM mass. For the light (heavy) case, the crossover point from the fast to the slow regime at low $M_0$ and high $n_{0}$ provides a lower (upper) bound on the DM mass.  As this point is a crossover between regimes, by continuity, we must find that the bounds are parametrically similar. This can indeed be seen in Table~\ref{tbl:DMMassRanges}. Excluding super-Planckian DM masses we again obtain a maximum $T_{\mathrm{RH}}$ in the heavy case, which, by continuity, is the same for the slow regime. From the slope of the mass contours, we see that we set an upper bound in the light case for low $M_{0}$ and $n_{0}$, deep in the fast region. Similarly, in the heavy case, we set a bound at the crossover for low $n_{0}$ and high $M_{0}$. 

If electroweak baryogenesis is to be accommodated in our model, then we further require that $T_{\mathrm{RH}}\gsim 200\,\mathrm{GeV}$. We include a contour for this reheat temperature in Fig.~\ref{fig:constraint_plots}. This contour is particularly constraining in the heavy case, yielding, for example, a revised range for the DM mass of $5.5\times10^{17}\,\mathrm{GeV}\lesssim\mdm\lesssim\mpl$ for a real scalar.

\begin{figure}
\centering
  \includegraphics[scale=0.47]{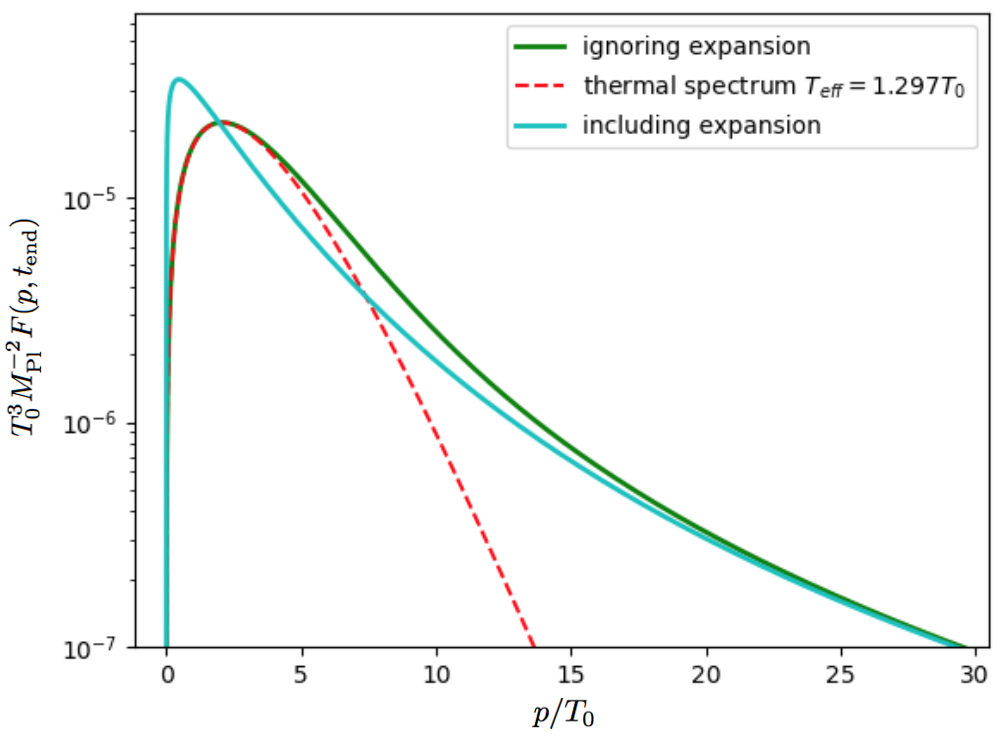}
\caption{\label{fig:free-streaming}
	DM momentum distribution function resulting at end of pBH decay. The green curve shows the distribution resulting from `fast' decay which exhibits a significant high energy tail compared to a thermal spectrum. The blue curve is an example of the distortion arising from a `slow' decay (same $T_0$ but larger $n_0$) when the interim expansion of the universe is significant.}

\end{figure}

{\bf Free-Streaming:}
The `light' DM mass case is significantly constrained by the free streaming of the ultra-relativistically-emitted DM particles: we now calculate the DM momentum distribution function $F(p,t)$ after the end of pBH decay. Here, $p$ denotes the magnitude of the 3-momentum.  The number of particles emitted by a BH per dof, per unit time, with energy in the range $(\omega,\omega+\text{d}\omega)$ is given by Eq.~\eqref{eq:emissionrate}, where the grey-body factor, $\Gamma_{i,s, \ell,h}(\omega)$, is related to the absorption cross-section of the BH by $\sigma_{i,s,h}(\omega) =  \pi \sum_l (2l+1) \Gamma_{i,s, \ell,h}(\omega)/\omega^{2}$.  
As the \emph{form} of $F(p,t)$ is dominantly determined by the ever-increasing BH temperature, $T(t)$, and the subsequent red-shifting of the DM momentum, and \emph{not} the grey-body factors to a first approximation, we may use $\sigma_{i,s,h}(\omega)= 27 \pi M^2/\mpl^4$ - the geometrical optics limit. Notice that, since this ignores small spin-dependent low-$\omega$ suppressions of the spectrum~\cite{Page:1976df} the result will be a slight underestimate of the total portion of relativistic (high $\omega$) particles.
The instantaneous distribution of emitted momenta is thus 
\begin{equation}
	\frac{\text{d}\dot{N}}{\text{d}p}(p,t) = \frac{27M(t)^2}{2\pi \mpl^4}  \frac{p^2}{e^{p/T(t)}\pm1}.
\end{equation}
Where we are justified in taking the ultrarelativistic limit $\mu \rightarrow 0$ for 'light' dark matter.
The final distribution $F(p,t)$ we are left with at some time $t$ after the BH has decayed away ($t_{\mathrm{end}}$) is a superposition of all the above instantaneous distributions, each red-shifted appropriately from its time of emission
\begin{equation}
	F(p,t) = \int_{t_0}^{t_{\mathrm{end}}} \text{d}\tau \: \frac{\text{d}\dot{N}}{\text{d}p}   \left(p\frac{R(t)}{R(\tau)},T(\tau)\right)  \frac{R(t)}{R(\tau)}.
	\label{eq:momentumdistribution}
\end{equation}
Here we are interested in finding this expression at the end of the BH decay, $F(p,t_\text{end})$.
In the `fast' regime, where the scale factor is approximately constant, this takes the form of $M^{2}_{\mathrm{Pl}} T_0^{-3} \tilde{F}(p/T_0)$, with $\tilde{F}$ dimensionless, and can be evaluated analytically in terms of poly-logarithms. We have numerically integrated the general case. In Fig.~\ref{fig:free-streaming} we show the exact solution for the bosonic `fast' case and a typical effect on the distribution of a `slower' decay. We find that at low energies the non red-shifted distribution can be fitted by an effective temperature $T_{\text{eff}} \approx 1.3 \; T_0$ (which is \emph{higher} than that of the SM thermal bath) and an additional pseudo-exponential high-energy tail.  Note that $T_{\text{eff}}$ does \emph{not} characterise the total energy density of the DM, or its number density, which is very far below that of the SM particles.
A `slower' decay distorts the lower energy spectrum towards zero as particles produced early are Hubble red-shifted during the BH lifetime. The tail is unaffected by this as its major contributors are produced in the final explosive moments of the BH life, a process that is intrinsically `fast'.

From the BH decay time onwards these distributions are simply redshifted by Hubble expansion as $p \sim 1/R$.  From this we deduce that, as the universe cools to some temperature $T_{\mathrm{SM}}$, the number of particles that are still relativistic is 
\begin{equation}
	\int_{p_{\min}}^{\infty} \mathrm{d}p\,F(p,t_{\mathrm{end}}), \quad  p_{\min} = \mu \frac{T_{\mathrm{RH}}}{T_{\mathrm{SM}}} \left(\frac{g_{*S}(T_\mathrm{RH})}{g_{*S}(T_\mathrm{SM})}\right)^{1/3}.
\end{equation}
Explicitly, in our two separate regimes (and taking $T_{\mathrm{SM}} = 1\,\mathrm{keV}$, see below) we have 

\begin{equation}
	p_{\min} \approx  \left(\frac{g_{*S}(T_\mathrm{RH})}{3.91}\right)^{1/3}  \frac{e_{\mathrm{T}}}{f_s g_s} T_0\left\{
\begin{matrix}
0.0285 & \text{`fast'}\\
0.0240 & \text{`slow'}
\end{matrix}
\right. ,
\end{equation}
where the spin dependence is made manifest. By inspecting Table~\ref{tbl:GreyBodyFactors} and Fig.~\ref{fig:free-streaming} it is clear that the lower the spin, the higher the proportion of relativistic particles.

\begin{figure*}[t]
	\begin{minipage}{\columnwidth}
		\includegraphics[width=0.47\columnwidth]{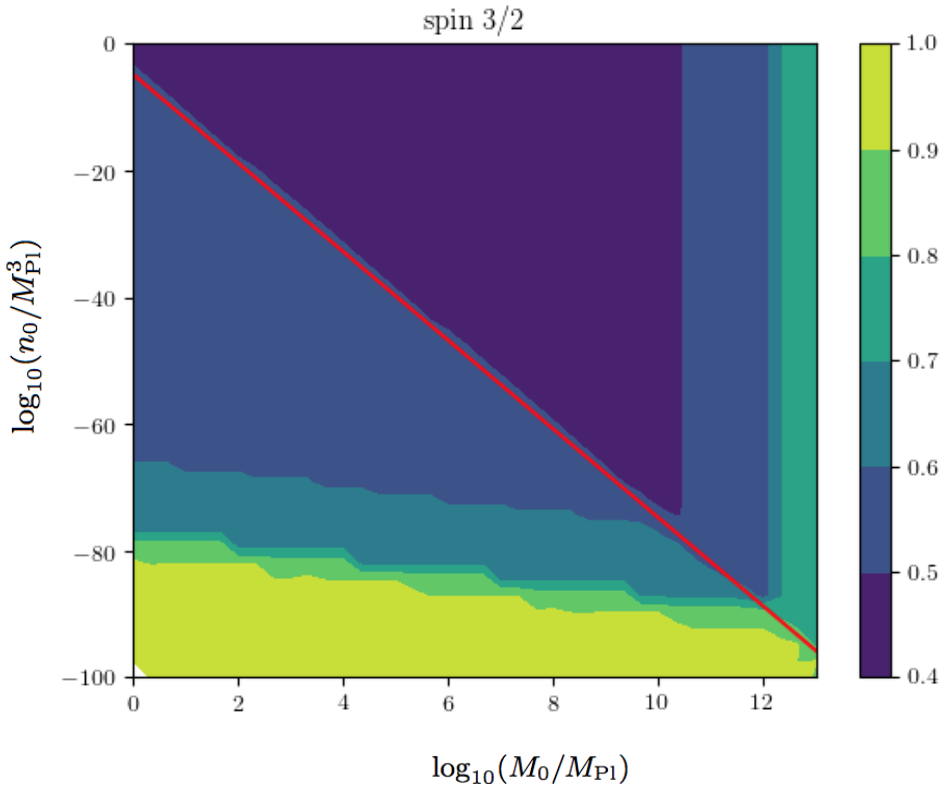}
		\hspace{0.5cm}
		\includegraphics[width=0.47\columnwidth]{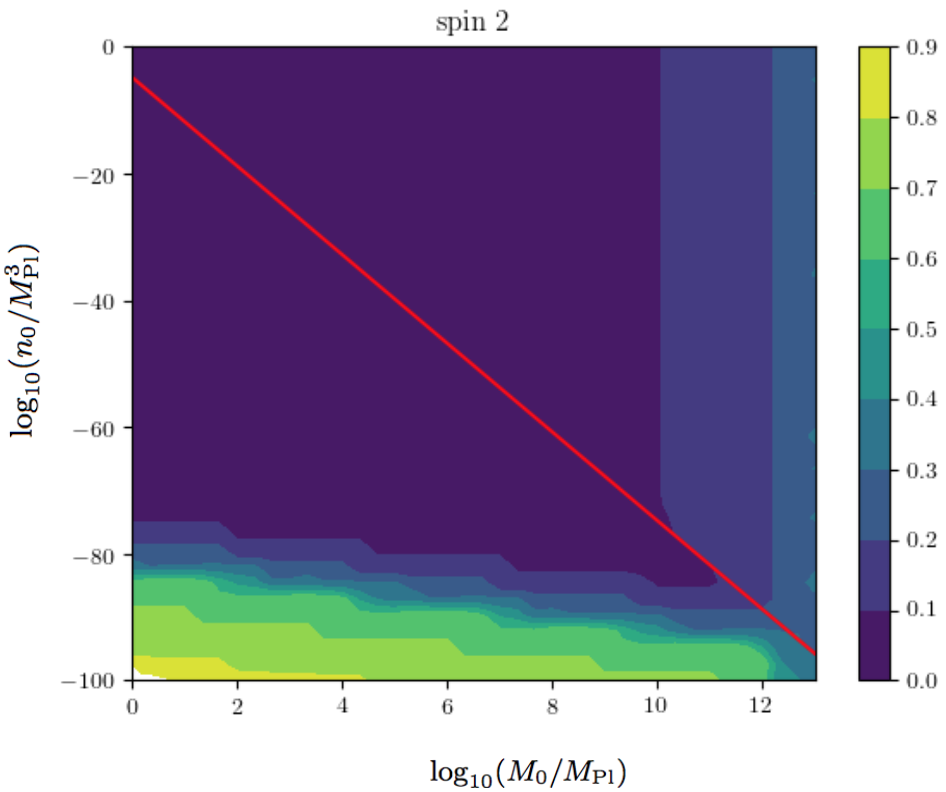}
	\end{minipage}
	\caption{
Free-streaming constraints in the `light' DM case for spin 3/2 and 2 (left and right panels), where
colour shading shows fraction of DM particles that are still relativistic at $T_{\mathrm{SM}}=1\,\mathrm{keV}$, and we have at every point imposed a `light' solution
DM mass such that the correct $\Omega_{\mathrm{DM}} h^2$ is reproduced.  Note the differing colour scales in the two
cases with the spin 3/2 case having more than $\sim 40\%$ of particles relativistic over the entire plane, while the spin 2 case has substantial regions where
less than $\sim 10\%$ of DM particles are relativistic. Red line ($B_0 = 1$) marks the boundary between the `fast' and `slow' regimes.}
	\label{fig:freestreaming_plots}
\end{figure*}

It is possible to numerically evaluate the above integral expression and, as a rough-and-ready criterion for successful structure formation we impose that when $T_{\mathrm{SM}} = 1\,\mathrm{keV}$ (at which stage the horizon mass is $\sim 10^9 M_\odot$) less than 10\% of the DM is relativistic (see, e.g. \cite{Hisano:2000dz} for a relevant discussion of free streaming constraints).  We find that for the `light' DM case almost the entire
parameter space is excluded. In particular we find that for the `light' DM mass solution the distribution of DM momenta at $T_{\mathrm{SM}} = 1\,\mathrm{keV}$ is almost
entirely relativistic for spins $s\leq 1$,  and so this regime is definitively excluded for these spins (at least for the basic DM genesis mechanism we discuss in this Section).  For $s=3/2$ we find that between roughly 45\% and 90\% of the DM is still relativistic depending on where in the allowed $M_0$-$n_0$ plane we are situated, and so this case is marginal at very best given the criterion for successful structure formation
that we have employed~\footnote{For earlier work on the spin 3/2 case, see~\cite{Khlopov:2004tn}.}.  In the case of `light' $s=2$ DM we find that there are substantial regions where 10\% or less of the DM is relativistic at $T_{SM} = 1\,\mathrm{keV}$,
so this case naively survives and provides possibly observable and even useful modifications to the galaxy structure power spectrum. 
The fraction of DM particles that are still relativistic at $T_{\mathrm{SM}}=1$keV in these two cases, $s=3/2, 2$ are shown in Fig.~\ref{fig:freestreaming_plots}.
We caution the reader, however, that a dedicated numerical simulation is required, including sub-leading effects such as the additional spectral distortions due to grey-body factors, to determine precisely the allowed parameter regime in the `light' case for spin $3/2$ and $2$.  We will return to a more detailed analysis of free-streaming in our mechanism in a future work.

In contrast, for the `heavy' DM mass solution, we find for all spins that 10\% or less of the DM is still relativistic at $T_{\mathrm{SM}} = 1\,\mathrm{keV}$, and thus
we have in this case a successful new mechanism of DM production in both the `slow' and `fast' regimes of BH decay.  

{\bf Dark Radiation:}
The change, compared to the SM, in the effective number of relativistic dof contributing to the energy density at the surface of last scattering is bounded from PLANCK2015 data~\cite{Ade:2015xua} by $\Delta N_{\mathrm{eff}}<0.1\pm0.2$ in terms of the number of effective neutrinos types (with $N_{\mathrm{eff,SM}} \simeq 3.046$). 
An unavoidable contribution to dark radiation comes from gravitons emitted from the BHs at a rate which is only suppressed compared to SM radiation by the smaller grey-body power coefficient, $e_2\simeq 1.92 \times 10^{-6} \ll e_{\mathrm{T},\mathrm{SM}} \simeq 4.38\times 10^{-3}$ (for related earlier work see~\cite{Dolgov:2011cq}).  Specifically at the end of pBH decay
$\Delta \rho_{\mathrm{grav}}/\rho_{\mathrm{rad}} = 2 e_2 / e_{\mathrm{T},\mathrm{SM}}\simeq 8.77\times 10^{-4}$, and this translates to $\Delta N_{\mathrm{eff}}$ by using the appropriate changes in the number of effective SM relativistic dof from $T_{\mathrm{RH}}$ down. 
In our case we find that the additional $\Delta N_{\mathrm{eff}}$ due to gravitons depends upon the reheat temperature: for $T_{\mathrm{RH}}\sim3\,\mathrm{MeV}$, $\Delta N_{\mathrm{eff,grav}}\simeq 5.39\times10^{-3}$ while for $T_{\mathrm{RH}}\sim1\,\mathrm{TeV}$, $\Delta N_{\mathrm{eff,grav}}\simeq 2.53\times10^{-4}$.  As $\Delta N_{\mathrm{eff,grav}}$ further decreases for increasing $T_{\mathrm{RH}}$, this does not
place a bound on our mechanism. Nevertheless, the larger figure may provide an accessible target for planned future high-precision determinations of  $N_{\mathrm{eff}}$.  In Section~IV we consider motivated variations of our basic model with additional dark radiation contributions.

{\bf Density Perturbations:}
Limits on the spectrum and form of density perturbations provide another potential source of constraint.  We have here taken as a basic hypothesis that on moderate-to-large scales (but not on small scales as we soon quantify) the initial BH number density, and thus the initial energy density, $\rho_{\mathrm{BH}}$, inherits the spectrum of Gaussian, almost scale-invariant density perturbations that are the result of, say, an early epoch of inflation.  (We have nothing new here to say about the origin of these perturbations. Note that we do not necessarily require inflation:  any mechanism or dynamics in the very early Universe that gives the 
approximately scale-invariant spectrum of observed density perturbations is suitable as long as there is a population of micro pBHs produced/present 
at the end satisfying the conditions set out in the introduction, and discussed further in Section IV.)  One significant point is that because the evaporation of our
pBHs sources all forms of matter (both SM and dark), the perturbations at the end of pBH decay are naturally \emph{adiabatic} and not
isocurvature.  

Nevertheless, many production mechanisms for pBHs, for instance their production in ultra-relativistic bubble collisions following, say, a quantum phase transition
(e.g., of the type discussed in \cite{GarciaGarcia:2016xgv}), lead to substantial random variation in the number of pBHs produced per initial Hubble volume, 
and thus to a significant extra source of density perturbations on \emph{small} scales of order $H_0^{-1}$.   
To quantify the potential later effect of this on structure formation, consider BHs that are initially distributed at $t_0$ with density variations between initial Hubble volumes $V_{H_0}$.   This implies at later times $\delta\rho / \rho (t) \approx 1/\sqrt{N} = 1/\sqrt{K(t) n_0 V_{H_0}(t)}$, where $K$ is the number of $V_{H_0}$ patches within the horizon at $t$ and $N$ is therefore the total number of BHs within that horizon.  If we assume a Poisson distribution of BHs, this approximation is very accurate at later times when $K$ is very large.  Assuming matter domination during BH decay, and radiation domination thereafter, we require that $\delta\rho / \rho$ at matter-radiation equality be at most of order $10^{-6}$.  We find that this constraint is very weak and the the parameter regions thus excluded lie well within the section of the plane already forbidden by the requirement that $T_{\mathrm{RH}} \geq 3$ MeV.   Thus no substantial new perturbations
exist on length scales of interest for moderate-to-large-scale structure formation.

\section{Comments, Generalisations, and Future Work}

Finally, we make some comments, and briefly discuss some motivated variations of our basic mechanism, as well as directions for future work.

{\bf Sensitivity to Initial Conditions:}
The reader might be concerned that we have so far assumed that the initial condition is one of pure pBH matter domination. However, we have found that our DM yield prediction is insensitive to changes in this assumption over a wide range of parameter space.  In the fast BH decay regime, provided any initial SM radiation component satisfies $\rho_{\mathrm{rad,SM}}(0) \lsim 0.1 \rho_{\mathrm{BH}}(0)$ the predicted SM reheat temperature is unaffected up to
${\cal O}(10\%)$ corrections or less.  Regarding the DM yield, any primordial DM number density $\Delta n_{\mathrm{DM}}(0)$ is unimportant if it is substantially less than that produced by the pBH decay.  Given the expression for the total number of DM particles emitted, Eq.~\eqref{eq:particlenumber}, 
this translates into the requirement 
\begin{equation}
\frac{\Delta n_{\mathrm{DM}}(0)}{n_{0}} \ll 2\times 10^{-3} \left(\frac{\mpl^3}{n_{0}}\right)^{2/7} ,
\label{eq:DMnumbercondition}
\end{equation}
where for concreteness we have inserted typical values for scalar or spin-1/2 fermion DM in the `light' case, and used the condition $B_0\sim 1$
which allows the largest (most restrictive) fast-regime value of the initial pBH number density $n_0$ for a given value of $M_0$. Since over much of the allowed
parameter space $n_{0}/\mpl^3$ can be as small as $10^{-80}$ Eq.(\ref{eq:DMnumbercondition}) is not a particularly restrictive requirement.

The situation in the slow BH decay regime is even better, as any initial SM radiation component is now substantially red-shifted away during the long
period of BH decay.  Specifically, we find from our analytic slow-regime solution that, at end of pBH decay, any initial SM radiation has been diluted by a
factor
\begin{equation}
\frac{\Delta\rho_{\mathrm{rad,SM}}(t_{\mathrm{end}})}{\Delta\rho_{\mathrm{rad,SM}}(t_0)} \simeq \left(\frac{B_0}{0.928}\right)^{8/3} \ll 1
\end{equation}
as $B_0<1$ in the slow regime, and often $B_0\ll 1$.  This is to be compared to the SM radiation component produced by pBH emission, which is $\rho_{\mathrm{rad,SM}}(t_{\mathrm{end}}) \simeq 0.965 B_0^2  \rho_{\mathrm{BH}}(t_0)$.  Thus as long as
\begin{equation}
\rho_{\mathrm{rad,SM}}(t_0) \lsim 0.1 B_0^{-2/3} \rho_{\mathrm{BH}}(t_0)
\end{equation}
both the  SM reheat temperature and the DM yield are substantially unaffected, up to ${\cal O}(10\%)$ corrections or less, by the presence of the early SM radiation component.  Since, consistent with constraints, we find that in the slow regime $B_0^{-2/3}$ can be as large as
\begin{equation}
B_0^{-2/3}\sim 0.24 (\mpl/T_{\mathrm{RH}})^{5/9} \gg 1~,
\end{equation}
\emph{a dominant initial SM radiation component, as big as $\sim 10^{11}$ times larger than the initial pBH energy density can be present without affecting our results}.  (This ratio is for $T_{\mathrm{RH}} = 3\,\mathrm{MeV}$.  For $T_{\mathrm{RH}} = 1\,\mathrm{TeV}$, a ratio as large as $\sim 10^8$ is still allowed.)  It is clear that the slow region is particularly insensitive to initial conditions, and thus attractive in this regard.  In addition the region of parameter space
where there is of order one BH per initial horizon patch, and so possibly favoured by BH production mechanisms, is contained within the slow regime.

{\bf Other DM Interactions:}
Although our mechanism requires that the traditional freeze-out
and recent freeze-in mechanisms of DM production give sub-dominant effects, this doesn't rule out some small self- or DM-SM-interactions with possibly
important observational consequences.  We thus now briefly turn to the possibilities for detection of DM through non-gravitational signatures. 
 
One natural possibility is the existence of operators which allow the DM to slowly decay to the
SM sector.  (See, e.g., \cite{Ellis:1990nb,Kuzmin:1997jua,Birkel:1998nx,Chung:1998zb,Ziaeepour:2000rc,Barbot:2002gt,Chou:2003wx,Arvanitaki:2008hq} for a selection of earlier work on the topic of decaying heavy DM.)  This is well motivated if the stabilising symmetry of the DM is global, as, by general arguments, all theories incorporating gravity are widely believed to possess no exact global symmetries.  (It is not known if the strength of this violation is suppressed by some large mass scale, such as the Planck mass, or is non-perturbatively small in some dimensionless coupling.)  Since our DM, especially in the `heavy' case, is surprisingly massive, even exponentionally tiny symmetry-violating effects can lead to cosmologically relevant decay times.  As a rule of thumb, lifetimes of order $10^{26}-10^{27}\,\mathrm{s}$ are on the allowed border for decaying DM, depending on exact details on the decay.  Since for our mechanism the DM is often very massive, the energy of the SM decay products can be very large, thus leading to ultra-high-energy cosmic rays.  If we parameterise the effective dimensionless coupling constant that violates the symmetry as $\lambda \sim \exp(-8\pi^2/g^2)$, as might be suitable for a non-perturbative violation, then for a DM mass of order the GUT scale, $\sim 10^{16}\,\mathrm{GeV}$, the $10^{26}\,\mathrm{s}$ lifetime constraint translates into $g<0.8$, a not unreasonable number for a fundamental dimensionless coupling.  Of course, such a non-perturbatively determined lifetime is very sensitive to the underlying and unknown coupling $g$ so an observationally interesting lifetime is in no way predicted.  An alternative, and less sensitive, possibility exists for a DM mass in the region of $100\,\mathrm{TeV}$, for then dimension-six Planck-suppressed operators lead to a lifetime of order $10^{26}\,\mathrm{s}$ with only mild polynomial dependence on the mass.  

The DM can also have self-interactions as long as they don't lead to a freeze-out process.  For example, in the case of a single real massive scalar 
a quartic interaction will not allow physical, kinematically allowed number-changing interactions, but will lead to elastic scattering which can
modify the DM momentum distribution.  

For our heavier mass regions the DM can also have substantial interactions with SM matter.  This is because for DM mass $\mdm\gg T_{\mathrm{RH}}$ the
freeze-in process is exponentially suppressed \cite{Hall:2009bx} in addition to being down by small couplings, while any potential freeze-out reprocessing
via, say, self-annihilations of our DM into SM states is suppressed by the extremely tiny number density of DM that applies in the case
that $\mdm\gg1\,{\mathrm{TeV}}$.   Some of these issues have previously been discussed in the context of so-called WIMPzilla DM \cite{Kolb:1998ki,Chung:1998zb}
but, in the context of our production mechanism, deserve a more detailed analysis which we leave for a future work.

Finally, regarding the identity of the DM, the fact that our allowed `heavy' DM mass range extends all the way up to $\mpl$, and, moreover, is operative
for states of spin $s\geq 3/2$, motivates the intriguing possibility that either GUT, or even higher-spin string Regge excitations, if they are sufficiently stable (say due to a combination of $J^{\mathrm{PC}}$ conservation, kinematics, and maybe discrete gauge symmetry quantum numbers), might be the DM!

{\bf Additional Dark Radiation:}
If there are new very light dof in addition to the graviton and SM neutrinos then these will also be produced during the decay of the pBHs and will be
an additional dark radiation component beyond that of the gravitons.
A particularly well-motivated almost massless dof is the axion, whether the QCD axion \cite{Peccei:1977hh, Weinberg:1977ma, Wilczek:1977pj} or multiple axion-like particles as in the string axiverse scenario \cite{Svrcek:2006yi,Arvanitaki:2009fg}. If the number of light axions, $N_a$, is much less than the number of dof at $T_{\mathrm{RH}}$, then the effective change in the number of dof is: for $T_{\mathrm{RH}}\sim3\,\mathrm{MeV}$, $\Delta N_{\mathrm{eff}}=0.10\,N_{a}$; for $T_{\mathrm{RH}}\sim1\,\mathrm{TeV}$, $\Delta N_{\mathrm{eff}}=4.87\times10^{-3}N_{a}$. For $N_{a}\leq3$, both of these are consistent with Planck measurements -- however, for the lower reheat temperature, this could be discovered or ruled out by future experiments (see, e.g.,~\cite{Andre:2013nfa, Wu:2014hta}).

{\bf Opening the `Light' DM  Window:}
Another variation of our basic mechanism is to allow the number of dof in the SM sector to be much increased beyond the 106.75 of the SM -- for instance, due to TeV-scale (R-parity violating) supersymmetry or strong dynamics.  This situation can possibly resurrect some of the low-spin, $s\leq 1$ `light' DM parameter space from the severe free-streaming constraint.  The reason for this is two-fold: a drastic increase in SM dof at high energy leads, to first order, to substantial extra redshift of DM momentum by the time the SM temperature reaches $1\,\mathrm{keV}$, and second, less of the original BH mass is injected into DM -- the yield is inversely proportional to the now larger total emission coefficient $e_{\mathrm{T}}$.  Thus, all other things kept fixed, the DM must be heavier to satisfy the $\Omega_{\mathrm{DM}} h^2$ observation and so becomes non-relativistic earlier.

{\bf Early Matter Domination \& BH Mergers:}
The \emph{minimal} DM genesis mechanism that we have presented so far has an early period of matter domination due to the pBHs. (As explained above during
the discussion of insensitivity to initial conditions, it is possible that a much dominant SM radiation component is initially present, thus making the period of matter 
domination by pBHs very short.  Note, the period of matter domination in the fast regime is effectively automatically non-existent, due to the decay of the BHs into radiation within a Hubble time.)  It is therefore possible that density perturbations re-entering the horizon could gravitationally collapse into bound structures comprised of pBHs. The pBHs could then coalesce into larger BHs before 
they evaporate and so survive either down to the present epoch, or, more dangerously, decay post-BBN and lead to observable changes in BBN or CMBR predictions, or to constrained gamma ray or ultra-high-energy cosmic ray backgrounds.  In principle the general characteristics of this competition
between decay and coalescence is determined by the ratio of the Hawking evaporation time $\tau_{\mathrm{decay}}= M_0^3/3\mpl^4  e_{\mathrm{T}}$ to 
the coalescence time $\tau_{\mathrm{coalesce}} \simeq 10^{-2} \times \mpl^6 a^4 / M_0^3$ predicted by gravity wave emission (here $a$ is the BH binary
separation, and we take quasi-circular orbits).   Thus evaporation occurs before coalescence if  $M_0 \lsim 0.2 (\mpl a)^{2/3} \mpl$.  

To evaluate the evaporation vs coalescence condition we need to know the distribution function of orbital radii $a$.  It is natural to assume that typical values of $a$ are set by the average distance ${\bar d}$ between pBHs in our scenario, which is growing due to Hubble expansion until the binary forms and the BHs separate from the Hubble flow.
The maximally restrictive assumption is that some BH binaries form early when ${\bar d} \sim (3/4\pi n_0)^{1/3}$, leading to the condition $M_0/\mpl \lsim 0.15 (n_0/\mpl^3)^{-2/9}$.  This cuts off the upper right corner
of the slow region, and interestingly also requires more than one pBH per initial Hubble volume, but still allows a significant part of the slow regime parameter
space to survive, as well as all of the fast BH decay regime.   
However,  we have ignored some complicating issues.  First, rare close binaries may form with much shorter coalescence times.  Second, binaries with
orbits of very high eccentricity, $e\sim 1$, also have much shorter coalescence times $\tau_{\mathrm{coalesce}} \propto (1-e^2)^{7/2}$.  Thus our discussion here is highly provisional, though it seems clear that much, and likely all, of the fast regime parameter space, and possibly much of the slow regime, survive unaffected by BH mergers.  We comment, though, that such mergers would, if they are frequent enough,  substantially increase the gravitational dark radiation component beyond that calculated in Section~III.   Moreover the result of such a merger
is a spinning Kerr BH, and for such BHs the Hawking emission of higher-spin particle states is enhanced, further increasing the gravitational dark radiation (and also
possibly the DM yield in the case $s>0$).

Finally, we estimate the maximal mass of BH that could form during this early period of matter domination.  A rough bound in the slow regime can be placed by determining the mass within the horizon at the end of the BH life. (Parametrically this mass is the same as the mass within the horizon at the end of the slow decay epoch.) This mass is inversely proportional to the reheat temperature, and for the lowest allowed reheat temperature, $T_{\mathrm{RH}}\sim3\,\mathrm{MeV}$, we have a horizon mass $\sim10^{4}M_{\odot}$ which, though $\ll M_{\mathrm{dwarf}}$,  is a possibly intriguing number as regards the seeding of supermassive galactic BHs.  

{\bf Summary:}
We intend to return in future work to the physics of BH mergers in our scenario, as well as the other aforementioned topics, including a more precise
characterisation of the free-streaming bounds.  It is our belief, however, that our calculations and discussion here presented have made a convincing case
that the mechanism of DM production by Hawking evaporation is both viable and behaviourally rich, and highly worthy of
further exploration.

\section*{Acknowledgments}

We gratefully thank Isabel Garcia~Garcia and Kazunori Kohri for useful and stimulating discussions. OL and HT are supported by the Science and Technology Facilities Council (STFC). RPB is supported by a Clarendon Scholarship from the University of Oxford.

\bibliographystyle{JHEP}
\bibliography{DM_from_pBH.bib}

\end{document}